\documentclass[a4paper,fleqn,longmktitle]{cas-sc}

\usepackage[numbers]{natbib}

\def\tsc#1{\csdef{#1}{\textsc{\lowercase{#1}}\xspace}}
\tsc{WGM}
\tsc{QE}
\tsc{EP}
\tsc{PMS}
\tsc{BEC}
\tsc{DE}

\begin{document}
\let\WriteBookmarks\relax
\def\floatpagepagefraction{1}
\def\textpagefraction{.001}
\shorttitle{Leveraging social media news}

\title [mode = title]{On the shear flow instability in the red spot of Jupiter}                      



\author[]{Gutishvili G.}[]


\address[]{School of Physics, Free University of Tbilisi, 0183, Tbilisi, Georgia}

\author[]{Osmanov Z.}[style=chinese]

\credit{Data curation, Writing - Original draft preparation}


\cortext[cor1]{Corresponding author}
\begin{abstract}
In the present paper we study instability caused by the velocity shear flows of the great red spot of Jupiter.  For the purpose, we imply the Navier €"Stokes and the continuity equations, perform the linear analysis of the governing equations and numerically solve them. We have considered two different regimes: exponential and harmonic behaviour of wave vectors. It has been shown that both scenarios reveal unstable character of flows and the corresponding rate might be considerably high. For the case of harmonic time dependence of wave vectors we also found a beat like solution of sound waves.

\end{abstract}

\begin{keywords}
 Jupiter's great red spot \sep hydrodynamic instabilities \sep non-modal shear flows
\end{keywords}

\maketitle

\section{Introduction}

The Great Red Spot (GRS) of Jupiter has attracted attention since its discovery by Giovanni Cassini. Interest of researchers has significantly increased after measuring the decrease of the lengthscales of the 
GRS  (see Ref. \citep{change}). It has been shown that the GRS suddenly decreased in longitudinal extent faster than the average contraction rate. In particular, it was found that according to the $2014$ Hubble data the corresponding decrease has been estimated as to be approximately $1760$km only in $21$ months, exceeding the average rate of change $4$ times. A new era of exploration has been started by the Juno spacecraft, the NASA space probe launched in $2011$ and now orbiting the Jupiter. Juno's mission is dedicated to the study of the magnetosphere, gravity and composition of the planet (see Refs. \citep{maggrav}-\citep{composit}). 

In the context of the present paper, recently a very interesting work has been published (see Ref. \citep{dynamics}), where the authors have implied high-resolution images of the GRS obtained by the Juno, during its flyby of Jupiter on $2011$ July, $11$. In particular, the authors have analysed the data of cloud morphologies. It has been found that filaments with lengthscales of the order of $2000-10000$km circulate at speeds with the characteristic scale $120-140$m/s.

The overall dynamics of GRS must be strongly influenced by the observed velocity shear flows (SF), in certain cases leading to unstable character of the induced waves (see Ref. \citep{modal}). In particular, when implying the modal scheme, the Navier Stokes equation reduces to the following general form $\partial_t\xi+\left({\bf V_0\cdot\nabla}\right)\xi = M \xi$, where $\xi$ and $M$ represent respectively a physical quantity and a Hermitian operator belonging to the Hilbert space and ${\bf V_0}$ is the unperturbed velocity. In the framework of this approach the physical quantities are presented by the so-called normal modes characterised by certain time constants $\sim e^{-i\Omega t}$, with corresponding frequency $\Omega$.

The operator, $M$, becomes non-Hermitian in case of the inhomogeneous velocity SFs, when the physical quantities cannot be represented by the aforementioned anzatz $\sim e^{-i\Omega t}$ (see Ref. \citep{tatsuno}). If this is the case the field components nontrivially depend on time and consequently the problem reduces to the initial value problem. Therefore, the physical processes influenced by the nonmodal SFs might be characterised by another class of solutions (see Ref. \citep{tref}), exhibiting very interesting properties even in relatively simple Couette flows (see Ref. \citep{couette}). 

The nonmodal approach has been applied to a series of astrophysical problems. In particular, in (Ref. \citep{vazha}) it has been shown that by means of the nonmodal shear flows electron-positron plasmas might be efficiently driven. The influence of the nonmodal SFs on instabilities in MHD and electrostatic waves has been studied in (see Refs. \citep{chven,electrost}) and the interesting astrophysical consequences in the context of SF driven heating process has been discussed in (\citep{andro}-\citep{orp12}).

In the framework of the modal approach, the linear wave instabilities caused by the velocity shear has been examined in (Ref. \citep{shear1}). The authors have considered the linear stability problem numerically and it has been found that a deep vertical shear can induce amplified modes with timescales of the order of $1$hour. 

Unlike the standard scheme, in the nonmodal mechanism, the wavelength might exponentially decrease,  inevitably driving the exponential amplification of excited waves. Another interesting feature is the so-called parametric instability, which occurs for certain values of parameters, despite the harmonic (non-eponential) behaviour of the wavelength. The nonmodal velocity SFs might be very efficient in astrophysical tornados (Ref. \citep{chven}). On the other hand, the GRS is a strongly vortical structure and therefore, it is interesting to study a role of the nonmodal instabilities in the mentioned object.

The paper is organized in the following way. In section~2, we present a theoretical model of velocity SF nonmodal instability, in section~3, we obtain results and in section ~4 we summarise them.

\section{Main consideration}

In this section, we work out a mathematical model to study flow dynamics in the GRS of Jupiter. 
For this purpose we use the the Navier Stokes equation:
\begin{equation}
\label{euler} D_t{\bf V} = -\frac{1}{\rho}{\bf \nabla }P + \frac{\eta}{\rho}\Delta {\bf V}+{\bf g},
\end{equation}
and the continuity equation
\begin{equation}
\label{conn} D_t\rho + \rho \nabla \cdot {\bf  V}= 0,
\end{equation}
complemented by the equation of state $P=C{\rho}^{n}$, where $D_t \equiv \partial_t + ({\bf V} \cdot \nabla)$, ${\bf V}$ represents velocity of the flow, $\rho$ and $P$ are respectively density and pressure of the flow, $n$ is the polytropic index of gas, $\eta$ is its viscosity and ${\bf g}$ is the free fall acceleration. 

The vertical motions in general might affect the flow dynamics, but since this work is the first attempt of this kind, we consider the flow motion in a surface perpendicular to ${\bf g}$. Another approximation that we use in the paper is a negligible role of viscous terms. It is clear from Eq. (\ref{euler}) that for small lengthscales the viscosity might become important, provided that the corresponding scales are of the order of $l\sim \eta/(\rho V)$, where $\eta\approx \rho\lambda\sqrt{2RT/(\pi\mu)}$, $\lambda\approx\mu/(\sqrt{2}A_0N_A)$ is the mean free path of molecules, $\mu\approx 2.02$ g mol$^{-1}$ and $T\approx 112$K are respectively the average molar mass and mean temperature of the upper atmosphere of GRS (Ref. \citep{molar}), $\rho\approx 1.7\times 10^{-5}$g cm$^{-3}$ is the corresponding density and $A_0\approx 10^{-16}$ cm$^2$ is the mean cross section of a molecule. For the mentioned parameters, the viscosity is of the order of $3\times 10^{-4}$ poise. Then, by taking into account the typical flow velocity, $120-140$m/s (Ref. \citep{dynamics}), one can straightforwardly show that the viscous terms become significant for lengthscales less than $l_c\sim 1.4\times 10^{-4}$ cm, which is much less than the global kinematic scale of the GRS. Therefore, viscosity does not impose any significant conditions on flow dynamics of GRS on its large scales.

To explore the velocity SF nonmodal instability in the linear approximation, let us make first order expansion of physical variables 
\[ \rho = \rho_{0} + \rho^{'} \]
\[ {\bf V} = {\bf V_{0}} + {\bf V^{'}} \]
leading to the following set of linearised equations

\begin{equation}\label{eul}
\mathcal{D}_t{\bf V'} + ({\bf V'} \cdot \nabla){\bf V_0} =
-\frac{C_s^2}{\rho_0}{\bf \nabla \rho'},
\end{equation}

\begin{equation}\label{con}
\mathcal{D}_t\rho' + \rho_0 (\nabla \cdot {\bf V'})= 0,
\end{equation}
where $\mathcal{D}_t \equiv\partial_t + ({\bf V_0}\cdot \nabla)$
and $C_s = \sqrt{dP/d\rho}$ is the sound speed calculated for the background flow.

The present method is a local study of development of disturbances. Therefore, at any point of the surface $A(x_0,y_0)$ according to the standard scheme we expand an unperturbed velocity by a Taylor series up to the first order terms
\begin{equation}\label{velexpand}
{\bf V}={\bf V}(A)+\sum_{i=1}^2\frac{\partial{\bf V}(A)}{\partial
x_i}(x_i-x_{i0}),\end{equation}
where $i=1,2$, $x_i=(x,y)$ and henceforth for denoting the unperturbed velocity we use ${\bf V}$ instead of ${\bf V_0}$.  As it will be realised, the velocity inhomogeneity, characterised by the following shear matrix
\begin{equation}\label{S}
 {\bf S} = \left(\begin{array}{ccc} V_{x,x} & V_{x,y}\\
V_{y,x} & V_{y,y}\\ 
\end{array} \right )\equiv\left(\begin{array}{ccc} \Sigma & A_1 \\
A_2 & -\Sigma\\ 
\end{array} \right ),\end{equation}
where $V_{i,k}\equiv\partial V_i/\partial x_k$ (with $V_i(A)$ denoting the zeroth order components of velocity) are very crucial parameters for describing the SF driven instabilities.

Following the standard scheme, for physical quantities we use the anzatz (see Ref. \citep{electrost})

\begin{equation}\label{anzatz}
F(x,y,t)\equiv\hat{F}(t)e^{i\left(\phi_1-\phi_2\right)},\end{equation}
with
\begin{equation}\label{fi1}
\phi_1\equiv\sum_{i=1}^2{K_i}(t)x_i,\end{equation}
and
\begin{equation}\label{fi2}
\phi_2\equiv\sum_{i=1}^2V_i(A)\int{K_i}(t)dt,\end{equation}
where $K_i(t)$ are the corresponding components of a wave vector.

As it has already been mentioned in the introduction, we are looking for a certain class of instabilities, caused by the velocity SFs. In particular, it can be shown (see Ref. \citep{tsereteli}) that the convective derivative of $F$

\begin{equation}\label{conv2}
\mathcal{D}_t F=e^{i\left(\phi_1-\phi_2\right)}\partial_t\hat{F}(t)+ix\left(K_x^{(1)}+a_1K_x+b_1K_y\right)F+iy\left(K_y^{(1)}+a_2K_x+b_2K_y\right)F,\end{equation}
considerably simplifies for wave vectors satisfying the following equation
\begin{equation}\label{dk}
{\bf \partial_{t}K}+ {\bf S^T} \cdot {\bf K}=0,\end{equation}
where ${\bf S^T}$ is the corresponding transposed matrix.

By introducing the dimensionless quantities 
$\textbf{u} \equiv \textbf{V}'/C_s$, $d \equiv
-i\rho'/\rho_0$ (henceforth we omit the superscript and instead of $\hat{F}$ we write $F$), $\textbf{k}\equiv \textbf{K}/K_x(0)$, $\sigma\equiv \Sigma/(K_x(0)C_s)$, $a_{1,2}\equiv A_{1,2}/(K_x(0)C_s)$, the Navier Stokes equation, the continuity equation and Eq. (\ref{dk}) write as
\begin{equation}\label{v}
{\bf u}^{(1)}+{\bf s}\cdot{\bf u} = {\bf k} d,\end{equation}
\begin{equation}\label{cont}
d^{(1)}+{\bf k} \cdot {\bf u} = 0,\end{equation}
\begin{equation}\label{k}
\textbf{k}^{(1)} + {\bf s^T} \cdot {\bf k} = 0, \end{equation}
where $\Psi^{(1)}$ denotes derivative with respect to dimensionless time $tK_x(0)C_s$.

One of the interesting features of the non modal instabilities is that the wave vector components evolve in time. Generally speaking, one can distinguish two different regimes of behaviour. In particular, from Eq. (\ref{k}) one can obtain
\begin{equation}\label{k}
k_{x} ^{(2)} + k_{x} \Gamma^{2} = 0,\end{equation}
and
\begin{equation}\label{k}
k_{y} ^{(2)} + k_{y} \Gamma^{2} = 0,\end{equation}
where  $  \Gamma^{2} \equiv -\sigma^{2}-a_{2}a_{1} $. As it is clear, if $\Gamma^2>0$ the wave vectors exhibit the harmonic behaviour with time, whereas for $\Gamma^2<0$, the wave vector components exponentially increase in time. It is not surprising that in the latter case the physical quantities might reveal exponential instability. An interesting scenario takes place when despite the harmonic time dependence of $k_{x,y}$, the density and velocity perturbations might undergo an unstable behaviour.

\section{Discussion}

In this section we will consider typical flow parameters of the GRS and will discuss the corresponding numerical results, with a special emphasis on instabilities. 

In the framework of the model we assume that the velocity configuration has the following general form given in polar coordinates
\begin{equation}\label{flow}
\textbf{V} = \{V_r, V_{\theta}\}= \{0, \upsilon(r)\},
\end{equation}
where $r$ is the corresponding radial coordinate. By taking into account the aforementioned velocity field one can straightforwardly show that the shear matrix is given by (see Appendix)
\begin{equation}\label{S1}
 {\bf S} = \left(\begin{array}{ccc} 0 & -\omega\\
 \\
r\frac{\partial\omega}{\partial r}+\omega & \;0\
\\

\end{array} \right ),\end{equation}

\begin{figure}[]
	\includegraphics[scale=0.6]{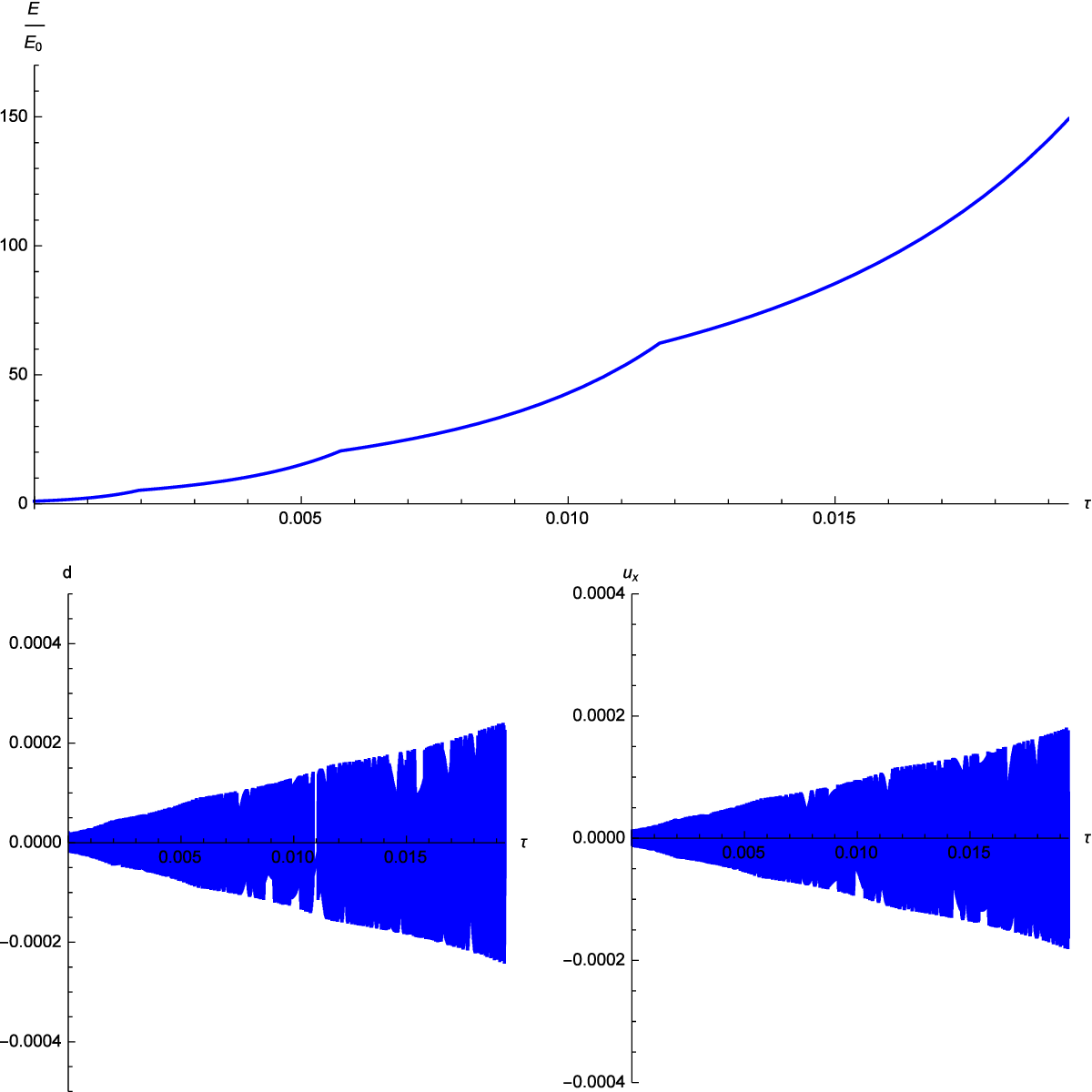}
	\caption{Here we demonstrate behaviour of $E$ 
normalised by its initial value (upper panel), density perturbation (bottom left panel) and $x$ component of velocity perturbation (all quantities on the graph are in dimensionless units). The set of parameters is: $L\approx 9\times 10^8$cm, $\upsilon = 140$m/s, $\partial\upsilon/\partial r\approx -3\times 10^{-5}$s$^{-1}$, $\lambda = L/1000$, $k_{x0}  = k_{y0} = 1$, $d_0 = 2\times 10^{-5}$, $u_{x0} = u_{y0} = 0$.}
	\label{Figure 1}
\end{figure}

\begin{figure}[]
	\includegraphics[scale=0.6]{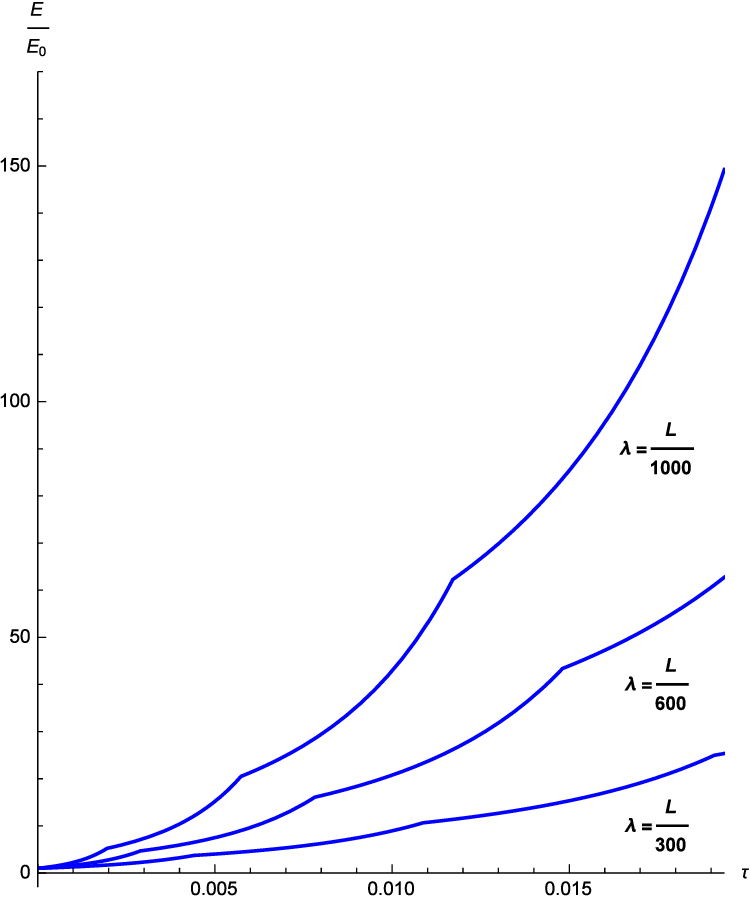}
	\caption{Here we plot the graphs of the normalised energy for different values of the wavelength $\lambda = \{L/1000; L/600; L/300\}$. The rest of the parameters is the same as on Fig. 1.}
	\label{Figure 2}
\end{figure}

where $\omega\equiv\upsilon/r$ represents the angular velocity of rotation.

As an example we consider the shear flow of a location approximately on the distance $L\approx 9\times 10^8$cm from the centre of GRS, where the velocity is of the order of $140$m/s and the linear velocity of rotation almost linearly increases with the radial coordinate $\partial\upsilon/\partial r\approx -3\times 10^{-5}$s$^{-1}$ (Ref. \citep{dynamics}). Hereafter we assume that the order of magnitude of the angular velocity is $\upsilon/L$. In Fig. 1 we show the dependence of dimensionless energy, $E\equiv {\bf u}^2/2+d^2/2$, normalised by its initial value (upper panel), density perturbation (bottom left panel) and $x$ component of velocity perturbation (all quantities are in dimensionless units). The set of parameters, apart from the above mentioned, is: $k_{x0}  = k_{y0} = 1$, $d_0 = 2\times 10^{-5}$, $u_{x0} = u_{y0} = 0$  (the subscript "0" indicates initial values of physical quantities) and the initial wavelength of the perturbation equals $\lambda = L/1000$. As it is clear from the plots, initially perturbed density (sound wave) drives the velocity perturbations and both reveal unstable behaviour. This is reasonable because $\Gamma^2<0$ and consequently $k_{x,y}(t)$ behave exponentially leading to the aforementioned unstable character of physical variables. Due to the SF nonmodal instability the initially induced sound waves pump energy from the mean flow and as a result the wave energy amplifies almost $150$ times during $\tau = 1.9\times 10^{-3}$. Hereafter time is normalised by the kinematic timescale of GRS, which is represented by the period of rotation - approximately $6$ Earth days. As it is clear from the plots, the instability is so efficient that approximately in $0.019\times 6\times 24\sim 2.7$ hours the waves amplify their energy by two orders of magnitude. We have already mentioned in the introduction that in the modal scheme the velocity shear induced instability has been considered in (Ref. \citep{shear1}) and the corresponding timescale is of the order of $1$ hour. However, one has to note that in the present study we examine velocity shear components in the GRS surface. Another difference is that in the framework of the nonmodal approach the wavelength significantly decreases, which in turn, might lead to heating or transition to a turbulence regime (but this is beyond the scope of the paper).

It is interesting to explore the dependence of instability on the initial value of $\lambda$. In Fig. 2 we show the normalised energy versus time for three different initial wavelengths a) $\lambda = L/1000$; b) $\lambda = L/600$; c) $\lambda = L/300$. The other parameters are the same as in Fig. 1. As we see from the plots, the less the values of the wavelength, the higher the corresponding energy of waves. 

\begin{figure}[]
	\includegraphics[scale=0.6]{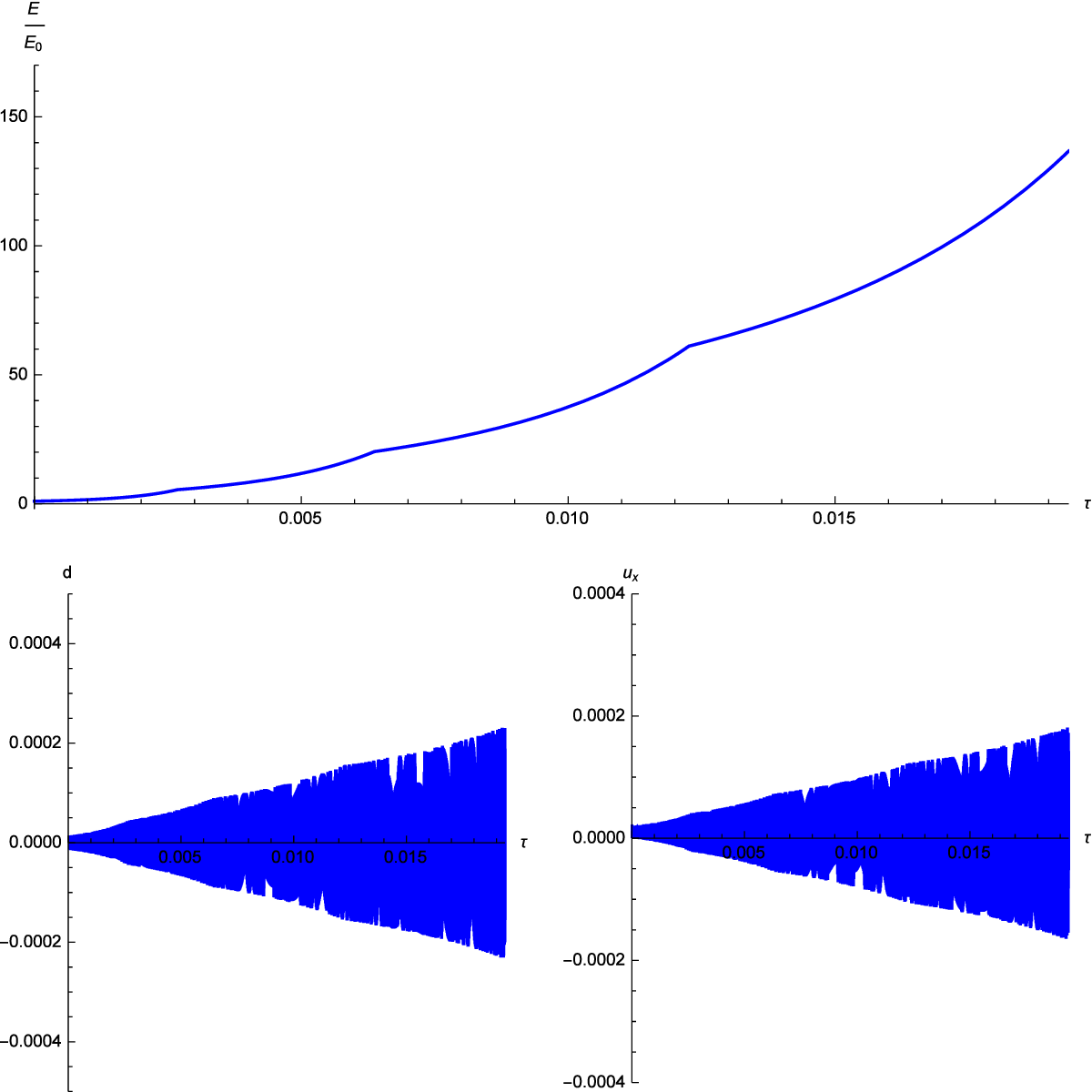}
	\caption{The behaviour of $E/E_0$, $u_x$ and $d$ versus time is shown. The set of parameters is the same as in Fig. 1, except $u_{x0} = 2\times 10^{-5}$, $d_{0} = 0$.}
	\label{Figure 3}
\end{figure}

In Fig. 1 it has been demonstrated that the initially excited sound waves drive the velocity perturbations. The opposite scenario is also possible. In particular, in Fig. 3 we show the time evolution of $E/E_0$, $u_x$ and $d$ respectively. The set of parameters is the same as in Fig. 1, except $u_{x0} = 2\times 10^{-5}$, $d_{0} = 0$. From the plots it is evident, that initially perturbed velocity component, in due course of time, induces the sound waves. Like the previous examples this particular case also reveals the high efficiency of the instability - the corresponding amplified energy of the wave exceeds the initial wave energy almost $150$ times indicating extremely high efficiency of the process.

\begin{figure}[]
	\includegraphics[scale=0.6]{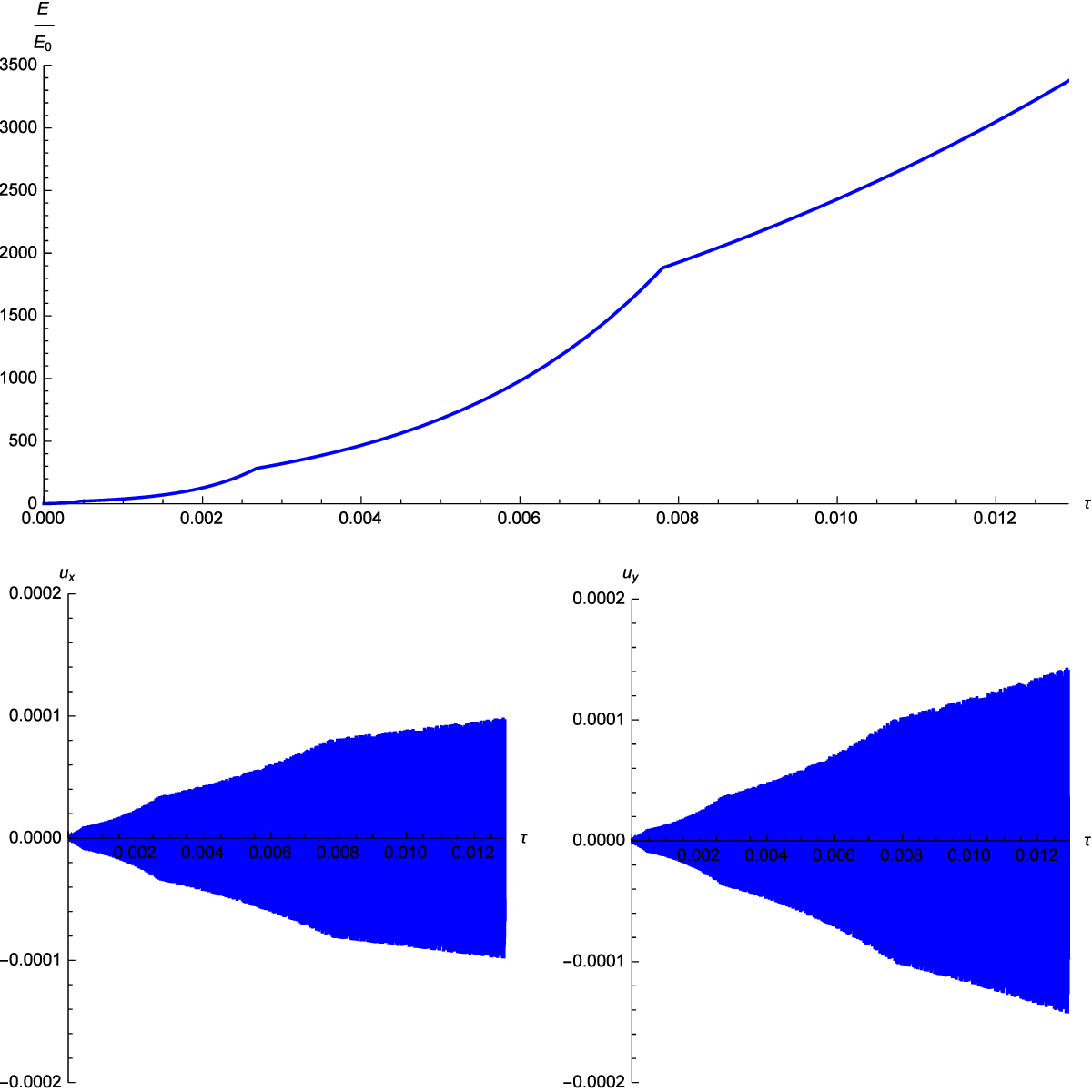}
	\caption{Here we plot the graphs of $E/E_0(\tau)$,  $u_{x}(\tau)$ and $u_{x}(\tau)$. The set of parameters is: $L\approx 6\times 10^8$cm, $\upsilon = 140$m/s, $\partial\upsilon/\partial r\approx 3\times 10^{-5}$s$^{-1}$, $\lambda = L/1000$, $k_{x0}  = k_{y0} = 1$, $d_0 = 3\times 10^{-6}$, $u_{x0} = u_{y0} = 0$.}
	\label{Figure 4}
\end{figure}
In the aforementioned examples $\Gamma^2<0$, the wave vector components behave exponentially, leading to the efficient instability. Unlike this scenario, there is a possibility that the instability occurs even for the flow satisfying $\Gamma^2>0$. In particular, inside the GRS on the location, $L\sim 6\times 10^8$cm, the velocity gradient is positive, $\partial\upsilon/\partial r\approx 3\times 10^{-5}$s$^{-1}$ and therefore $\Gamma^2>0$, leading to the harmonic behaviour of the wave vector components. In Fig. 4 the behaviour of $E/E_0$,  $u_{x}$ and $u_{x}$ versus time is shown. As it is clear from the plots, despite the condition $\Gamma^2>0$, the physical system undergoes SF instability resulting in the process of energy pumping from the mean flow to the excited waves. The peculiarity of the parametric instability is that it appears for a certain  interval of parameters. In particular, one can numerically check that the instability shown on Fig. 4 occurs only for the interval $1.05\times 10^{-6}<d_0<6\times 10^{-5}$.

\begin{figure}[]
	\includegraphics[scale=0.6]{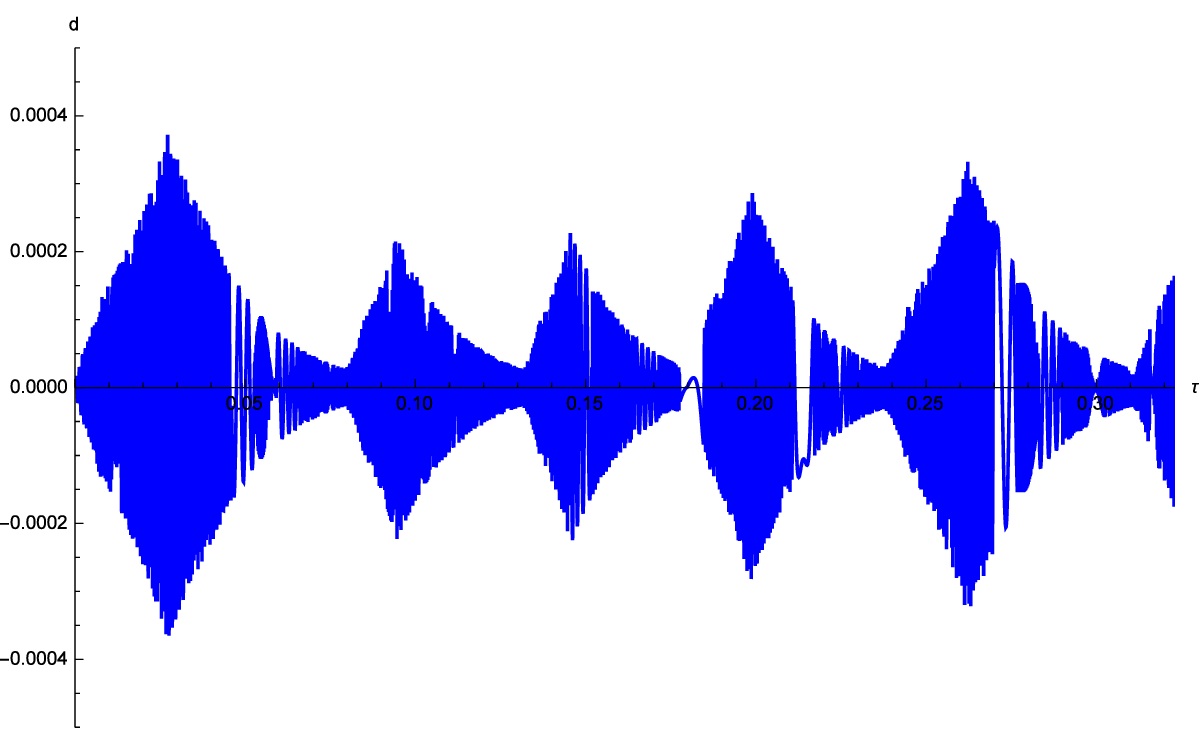}
	\caption{Here we demonstrate the time evolution of density. The set of parameters is the same as in Fig. 4 except $u_{y0} = 2\times 10^{-5}$ and $d_0 = 0$.}
	\label{Figure 5}
\end{figure}

Another interesting phenomenon which occurs for velocity SFs is the so-called beat-like behaviour of physical quantities. In Fig. 5 we demonstrate the dependence of  $d$ on time. The set of parameters is the same as in Fig. 4 except $u_{y0} = 2\times 10^{-5}$ and $d_0 = 0$. It is evident from the plot that for the mentioned parameters density reveals the beat-like behaviour.

Generally speaking, the SF nonmodal waves might significantly influence physical processes in flows. In particular, as it has been shown in the manuscript, under certain conditions the GRS flow may undergo efficient instability, when the wave vector components increase exponentially. As a result, the (sound, or vortex) waves efficiently pump energy from the mean flow. One of the direct consequences of this process might be an anomalous heating of an GRS ambient, because by decreasing the wavelength, the increasing role of viscosity might in principle terminate the amplification of the induced modes, finally converting energy to heat. Another interesting consequence of the nonmodal instability is the transition to turbulence, occuring for small lengthscales, which in turn, is naturally provided by the exponential character of $k_{x,y}(\tau)$. But this is very nonlinear process and like the previous phenomenon (heating) is beyond the intended scope of the present work.

\section{Summery}
In this paper we have considered the nonmodal study of velocity SFs. For this purpose we have examined the Navier Stokes equation, the continuity equation and the Poisson equation respectively. After linearising the aforementioned set of equations and applying the Fourier transform, we arrive at the set of ordinary differential equations.

Two general examples have been examined: the exponential and harmonic behaviour of the wave vector components. In the former case the physical system undergoes the SF instability. For a certain set of parameters, typical for the GRS it has been shown that initially excited sound waves efficiently convert energy from the mean flow and the corresponding energy amplifies by the factor of $150$ in approximately $2.7$ hours. Also it has been shown that the sound waves efficiently drive velocity perturbations. For the situation when initially only the velocity components are excited, the sound waves are driven as well and the amplification factor is of the same order of magnitude as in the previous case. 

By examining the condition $\Gamma^2>0$, leading to the harmonic behaviour of $k_{x,y}(\tau)$ (which takes place inside GRS) it has been shown that only for a relatively narrow range of initial values of density the instability occurs. In the framework of the same condition beat-like solutions have been found. This is particularly interesting, because such a behaviour might be a good fingerprint to detect the nonmodal SF instabilities in the atmosphere of GRS.

\section*{Acknowledgments}

The research was supported by the Shota Rustaveli National Science Foundation grant (DI-2016-14) and partially by the grant (FR17-587). The research of G.G. was supported by the Knowledge Foundation at the Free University of Tbilisi.

\appendix

\section{Appendices}
Below we give the derivation of Eq. (\ref{S1}). In the standard polar coordinates, the velocity components 
are given by

$$V_x = -r\omega(r)\sin\theta,
$$
$$V_y = r\omega(r)\cos\theta,
$$%
where $\omega(r) = \upsilon(r)/r$ is the local angular velocity of rotation and $\theta$ denotes the polar angle. By using the identities $\cos\theta = x/r$ and $\sin\theta = y/r$ the aforementioned equations reduce to
$$V_x = -y\omega(r),
$$%
$$V_y = x\omega(r).
$$%
We assume cylindrical symmetry, therefore one can consider a particular point $y = 0$, $x = r$ which does not violate generality of the approach. Then by taking into account the relations 
$$\frac{\partial r}{\partial x} = \frac{\partial }{\partial x}\left(\sqrt{x^2+y^2}\right)=\frac{1}{2\sqrt{x^2+y^2}}
\frac{\partial}{\partial x}\left( x^2+y^2\right)=$$
$$=\frac{2x}{2\sqrt{x^2+y^2}}=\frac{x}{r}.
$$%
and
$$\frac{\partial r}{\partial y} = \frac{\partial }{\partial y}\left(\sqrt{x^2+y^2}\right)=\frac{1}{2\sqrt{x^2+y^2}}
\frac{\partial}{\partial y}\left( x^2+y^2\right)=$$
$$=\frac{2y}{2\sqrt{x^2+y^2}}=\frac{y}{r}.
$$%
for the matrix elements we obtain
$$V_{xy}=\frac{\partial V_x}{\partial y} = -\frac{\partial}{\partial y}\left(y\omega(r)\right)=-\omega(r)-y\frac{\partial\omega}{\partial y}|_{y=0}=-\omega.
$$%
$$V_{yx}=\frac{\partial V_y}{\partial x} = \frac{\partial}{\partial x}\left(x\omega(r)\right)=\omega(r)+x\frac{\partial\omega}{\partial  x}|_{x=r}=\omega+r\frac{\partial\omega}{\partial r}\frac{\partial r}{\partial x}=$$
$$=\omega+\frac{x^2}{r}\frac{\partial\omega}{\partial r}|_{x=r}=\omega+r\frac{\partial\omega}{\partial r}.
$$%

\end{document}